\documentclass[superscriptaddress,twocolumn,showpacs,amsmath,amssymb]{revtex4}

\usepackage{graphicx}

\renewcommand{\vec}[1]{\mbox{\boldmath $#1$}}

\begin{document}

\title{Strong dineutron correlation in $^8$He and $^{18}$C}

\author{K. Hagino}
\affiliation{ 
Department of Physics, Tohoku University, Sendai, 980-8578,  Japan} 

\author{N. Takahashi}
\affiliation{ 
Department of Physics, Tohoku University, Sendai, 980-8578,  Japan} 

\author{H. Sagawa}
\affiliation{
Center for Mathematical Sciences,  University of Aizu, 
Aizu-Wakamatsu, Fukushima 965-8560,  Japan}


\begin{abstract}
We study the spatial structure of four valence neutrons in 
the ground state of $^8$He and $^{18}$C nuclei using a core+4$n$ model. 
For this purpose, we employ a density-dependent contact interaction 
among the valence neutrons, and solve the five-body Hamiltonian 
in the Hartree-Fock-Bogoliubov (HFB) approximation. 
We show that two neutrons with the coupled spin of $S$=0 
exhibit a strong dineutron correlation around the surface of 
these nuclei, whereas 
the correlation between the two dineutrons is much weaker. 
Our calculation indicates that the probability of the (1p$_{3/2})^4$ 
and [(1p$_{3/2})^2$ (p$_{1/2})^2$] configurations in the ground state
wave function of $^8$He nucleus is 34.9\% and 23.7\%, respectively. 
This is consistent with the recent experimental 
finding with the $^8$He($p,t)^6$He reaction, that is, the ground state 
wave function of $^8$He deviates significantly from 
the pure (1p$_{3/2})^4$ structure. 
\end{abstract}


\maketitle

\section{Introduction}

It has been well recognized that the pairing correlation 
and couplings to the continuum spectra play an essential role 
in weakly bound nuclei \cite{BE91,DNW96}. 
Although the dineutron structure 
as a consequence of the pairing correlation 
has been suggested for some time in 
$^{11}$Li and $^6$He nuclei\cite{BE91,ZDF93}, 
it is only recently that 
a strong indication of  
its existence has been obtained experimentally in the 
Coulomb dissociation of $^{11}$Li \cite{N06}. 
The new measurement has stimulated lots of theoretical 
discussions on the dineutron correlation, not only in 
the 2$n$ halo nuclei, $^{11}$Li and $^6$He
\cite{BBBCV01,HS05,HSCP07,BH07}, 
but also 
in medium-heavy neutron-rich nuclei \cite{MMS05,PSS07} 
as well as in infinite neutron matter \cite{M06,MSH07}. 

In Ref. \cite{HSCP07}, we have studied 
the behaviour of valence neutrons in $^{11}$Li
at various positions from the center to the surface of the nucleus.
We have found that i) the two-neutron wave function
oscillates near the center whereas it becomes similar to
that for a bound state around the nuclear surface,
and ii) the local pair coherence length has a
well pronounced minimum around the nuclear surface.
This 
result clearly indicates that a strong di-neutron correlation 
between the valence neutrons is present 
on the surface of the nucleus. 

An important next question is 
how the spatial structure of valence neutrons 
evolves 
from that in the 2$n$-halo nucleus, $^{11}$Li, 
when there are more numbers of neutrons. 
Although Refs. \cite{MMS05,PSS07} have partially addressed this
question by studying a two-particle density 
for medium-heavy neutron-rich nuclei, one would also need 
to explore a four-particle density, or many-particle density in
general, 
in order to shed light on 
the ground state properties of neutron-skin nuclei. 

For this purpose, $^8$He makes the most suitable nucleus to study. 
$^8$He is expected to have the $\alpha$+4$n$ 
structure \cite{THKSST92,SI88,VSO94,ZKS94,NVC01,AD06,KE07,IIAA08}, and thus 
provides a bridge between the 2$n$-halo nuclei and heavier skin nuclei. 
We mention that the spatial structure of the four valence 
neutrons in $^8$He
has been discussed in Ref. \cite{ZKS94}, where 
the authors constructed the ground state wave function 
by assuming that the
four neutrons occupy the 1$p_{3/2}$ state in a harmonic oscillator
potential. 
However, this model is too simplistic, since 
it completely neglects 
the pairing correlation
and the continuum couplings. Notice that the mixing of many partial
wave components, especially those with different parities, is
essential in order to have a spatially compact dineutron structure 
\cite{MMS05,PSS07,CIMV84}. 
In fact, we do not see any indication of dineutron correlation in the
result of Ref. \cite{ZKS94} (see Fig. 2 of Ref. \cite{ZKS94}), 
despite that the dineutron structure is expected to be enhanced 
in many neutron-rich nuclei \cite{HSCP07,MMS05,PSS07}. 

The purpose of this paper is to reinvestigate the spatial structure of
the four valence neutrons in $^8$He by taking into account
consistently the pairing and the continuum effects. To this end, we
use the core+4$n$ model, and diagonalize the five-body Hamiltonian 
in the Hartree-Fock-Bogoliubov (HFB) approximation. 
We also study the $^{18}$C nucleus as another nucleus which is
expected to have the core+4$n$ structure\cite{SMA04,HY06,HS07,VM95}. 
We will demonstrate below that the pairing correlation leads to 
the strong dineutron structure in $^8$He and $^{18}$C, in contrast to the result
of Ref. \cite{ZKS94}. 

The paper is organized as follows. 
In Sec. II, we detail the HFB method based on the core+4$n$ model. 
In Sec. III, we apply the method to $^6$He, where the result of the
exact diagonalization of the three-body ($\alpha$+$n$+$n$) Hamiltonian 
has been obtained \cite{HS05,EBH99}. 
We compare the HFB result with the exact result, and discuss the
applicability of the HFB method for the study of the spatial 
structure of valence neutrons. 
In Sec. IV, we present the results for the 
$^8$He and $^{18}$C nuclei. We discuss the two- and four-particle
densities, as well as the probability of the single-particle
components in the ground state wave function. We then summarize
the paper in Sec. V. 

\section{HFB method for a core+4$n$ model}

\subsection{HFB equations}

In order to study the structure of the $^8$He and $^{18}$C nuclei, we
employ the core+4$n$ model, and consider the following Hamiltonian:
\begin{eqnarray}
H&=&
\sum_{i=1}^4\left[\frac{\vec{p}^2_i}{2m_N}+V_{nC}(r_i)\right]
+\sum_{i<j}\,v_{nn}(\vec{r}_i-\vec{r}_j)-T_{cm}, \\
&\sim&
\sum_{i=1}^4\left[\frac{\vec{p}^2_i}{2m_N}
\left(1-\frac{1}{A}\right)+V_{nC}(r_i)\right]
+\sum_{i<j}\,v_{nn}(\vec{r}_i-\vec{r}_j),  \nonumber \\
\\
&\equiv&
\sum_{i=1}^4\left[\frac{\vec{p}^2_i}{2m}+V_{nC}(r_i)\right]
+\sum_{i<j}\,v_{nn}(\vec{r}_i-\vec{r}_j). 
\label{H5body}
\end{eqnarray}
Here, $m_N$ is the nucleon mass, 
$A$ is the mass number of the nucleus, 
$V_{nC}$ is a potential between a 
valence neutron and the core nucleus, and $v_{nn}$ is the pairing
interaction among the valence neutrons. 
$T_{cm}$ is the kinetic energy for the center of mass motion of the 
whole nucleus. 
In Refs. \cite{HS05,EBH99}, the center of mass motion is treated
exactly by introducing the recoil kinetic energy of the core nucleus 
(see Ref. \cite{SI88} for the derivation of the recoil term). 
In this paper, we approximate the treatment by taking only the 
diagonal components in 
$T_{cm}$, as is often done in mean-field calculations 
\cite{BRRM00}
(notice that 
the off-diagonal components contribute only to the exchange part of
the mean-field potential). 
This leads to the renormalization of the nucleon mass, 
$m=A/(A-1)\cdot m_N$. 

Although the five-body Hamiltonian (\ref{H5body}) could be diagonalized
exactly {\it e.g.,} with the stochastic variational method
\cite{VSO94}, we seek an approximate solution using the
HFB method \cite{DNW96,RS80,DFT84,B00}. 
The ground state wave function in the HFB method is given by \cite{RS80}
\begin{equation}
|{\rm HFB}\rangle = \prod_k\hat{\beta}_k|0\rangle,
\label{HFBwf}
\end{equation}
where the quasi-particle operator $\hat{\beta}_k$ is given by
\begin{equation}
\hat{\beta}_k=\int d\vec{r}\sum_\sigma\left(U^*_k(\vec{r},\sigma)a_{\vec{r}\sigma}
+V^*_k(\vec{r},\sigma)a^\dagger_{\vec{r}\sigma}\right).
\end{equation}
In this equation, $\sigma=\pm1/2$ is the spin coordinate, 
$a^\dagger_{\vec{r}\sigma}$ is the creation operator of nucleons at 
the position $\vec{r}$ and $\sigma$, and 
$U_k$ and $V_k$ are the HFB quasi-particle wave functions. 
In this paper, we employ a density-dependent pairing interaction
\cite{BE91} for $v_{nn}$ given by
\begin{equation}
v_{nn}(\vec{r},\vec{r}')=V_0\,\left(1-\frac{\rho_t(\bar{\vec{r}})}{\rho_0}\right)
\delta(\vec{r}-\vec{r}'),
\end{equation}
where $\bar{\vec{r}}=(\vec{r}+\vec{r}')/2$ and 
$\rho_t(\vec{r})=\rho_C(\vec{r})+\rho_{v}(\vec{r})$ is the total
density, $\rho_C$ and $\rho_{v}$ being the density of the core nucleus
and the valence neutrons, respectively. 
For this interaction, the expectation value of the Hamiltonian
(\ref{H5body}) with the HFB state (\ref{HFBwf}) reads \cite{DNW96}
\begin{eqnarray}
E&=&\langle {\rm HFB}|H|{\rm HFB}\rangle, \\
&=&
\int
d\vec{r}\left(\frac{\hbar^2}{2m}\tau(\vec{r})+V_{nC}(\vec{r})\rho_{v}(\vec{r})\right)
\nonumber \\
&&+\frac{V_0}{4}
\int d\vec{r}
\left(1-\frac{\rho_t(\vec{r})}{\rho_0}\right)
(\rho_{v}(\vec{r})^2+\tilde{\rho}_{v}(\vec{r})^2),
\label{EHFB}
\end{eqnarray}
where the kinetic energy density $\tau(\vec{r})$, the particle density
$\rho_{v}(\vec{r})$, and the pairing density 
$\tilde{\rho}_{v}(\vec{r})$ are given by 
\begin{eqnarray}
\tau(\vec{r})&=&\sum_k\sum_\sigma |\nabla V_k(\vec{r},\sigma)|^2, 
\label{tau}
\\
\rho_{v}(\vec{r})&=&\sum_k\sum_\sigma |V_k(\vec{r},\sigma)|^2, 
\label{rho}
\\
\tilde{\rho}_{v}(\vec{r})&=&-\sum_k\sum_\sigma
V_k(\vec{r},\sigma)U^*_k(\vec{r},\sigma),
\label{rhopair}
\end{eqnarray}
respectively. 
In deriving Eq. (\ref{EHFB}), we have used the properties of
time-reversal symmetry \cite{DNW96}. 

The equations for the HFB wave functions 
$V_k$ and $U_k$ are obtained by taking the variation of the energy
expectation value (\ref{EHFB}) with respect to the particle and the
pairing densities. 
This leads to the HFB equations in the coordinate space representation 
\cite{DNW96,DFT84,B00}, 
\begin{equation}
\left(
\begin{array}{cc}
\hat{h}-\lambda&\Delta(r) \\
\Delta(r)&-\hat{h}+\lambda
\end{array}
\right)
\left(
\begin{array}{c}
U_k(\vec{r},\sigma)\\
V_k(\vec{r},\sigma)
\end{array}
\right)
=E_k
\left(
\begin{array}{c}
U_k(\vec{r},\sigma)\\
V_k(\vec{r},\sigma)
\end{array}
\right),
\label{HFB}
\end{equation}
where $\lambda$ is the Fermi energy. The mean-field Hamiltonian
$\hat{h}$ is given by 
\begin{eqnarray}
\hat{h}&=&\frac{\delta E}{\delta \rho_{v}}, \\
&=&-\frac{\hbar^2}{2m}\nabla^2+V_{nC}(\vec{r}) 
+\frac{V_0}{2}
\left(1-\frac{\rho_t(\vec{r})}{\rho_0}\right)\rho_{v}(\vec{r})
\nonumber \\
&&
-\frac{1}{4}\,V_0\,
\frac{\rho_{v}(\vec{r})^2+\tilde{\rho}_{v}(\vec{r})^2}{\rho_0},
\label{hmf}
\end{eqnarray}
while the pairing potential $\Delta(r)$ is given by 
\begin{eqnarray}
\Delta(r)&=&\frac{\delta E}{\delta \tilde{\rho}_{v}}, \\
&=&
\frac{V_0}{2}\,
\left(1-\frac{\rho_t(\vec{r})}{\rho_0}\right)\tilde{\rho}_{v}(\vec{r}). 
\label{vpair}
\end{eqnarray}

We solve the HFB equations (\ref{HFB}) self-consistently 
by expanding the HFB wave
functions on the eigen functions of the mean-field Hamiltonian
$\hat{h}$ \cite{HS05b}. In doing so, 
we respect the Pauli principle and explicitly exclude those
states which are occupied by the neutrons in the core nucleus. 
Notice that the HFB method could be applied to light neutron-rich
nuclei without introducing the core
nucleus \cite{MR96,SDNPD03}.  
We nevertheless treat only the valence neutrons explicitly, since it
is not straightforward to 
separate between the core and the valence parts
from the HFB ground state wave function (\ref{HFBwf}). 

\subsection{Two- and four-particle densities}

In order to discuss the spatial structure of the valence neutrons, we 
compute the two- and four-particle densities using the solution of the
HFB equations. Using Wick's theorem, the two-particle density can be
expressed as \cite{MMS05}
\begin{eqnarray}
\rho_2(\vec{r}\sigma,\vec{r}'\sigma')&=& 
\langle {\rm HFB}|a^\dagger_{\vec{r}\sigma}a^\dagger_{\vec{r}'\sigma'}
a_{\vec{r}'\sigma'}a_{\vec{r}\sigma}|{\rm HFB}\rangle, \\
&=&
|\tilde{\rho}_v(\vec{r}\sigma,\vec{r}'\tilde{\sigma}')|^2
-|\rho_v(\vec{r}\sigma,\vec{r}'\sigma')|^2 \nonumber \\
&& +\rho_v(\vec{r}\sigma)\rho_v(\vec{r}'\sigma'),
\label{rho_2}
\end{eqnarray}
where $\tilde{\sigma}=-\sigma$, and the off-diagonal components of the
densities are given as (see Eqs. (\ref{rho}) and (\ref{rhopair})), 
\begin{eqnarray}
\rho_{v}(\vec{r}\sigma,\vec{r}'\sigma')&=&\sum_k
V_k(\vec{r},\sigma)V^*_k(\vec{r}',\sigma'), 
\label{rhomat}
\\
\tilde{\rho}_{v}(\vec{r}\sigma,\vec{r}'\sigma')
&=&-\sum_k V_k(\vec{r},\sigma)U^*_k(\vec{r}',\sigma'). 
\end{eqnarray}

In order to evaluate the four-particle density, 
\begin{eqnarray}
&&\rho_4(x_1,x_2,x_3,x_4) \nonumber \\
&&=\langle {\rm HFB}|a^\dagger_{x_1}a^\dagger_{x_2}a^\dagger_{x_3}a^\dagger_{x_4}
a_{x_4}a_{x_3}a_{x_2}a_{x_1}
|{\rm HFB}\rangle, 
\label{rho4}
\end{eqnarray}
where $x=(\vec{r},\sigma)$, 
we find it useful to express the HFB ground state wave
function, Eq. (\ref{HFBwf}), using the canonical basis. 
The canonical basis function $\psi$ is the eigenfunction of the density matrix 
(\ref{rhomat}) and satisfies \cite{DNW96,RS80}
\begin{equation}
\sum_{\sigma'} \int d\vec{r}'
\rho_{v}(\vec{r}\sigma,\vec{r}'\sigma')\,\psi_p(\vec{r}',\sigma')
=v_p^2\,\psi_p(\vec{r},\sigma). 
\end{equation}
In this paper, we construct the canonical basis by expanding 
$\psi_p$ on the HF basis, as is done for the HFB wave functions (see
the previous subsection). 
Using the canonical basis, the HFB ground state wave function is given
in the BCS form as \cite{DNW96,RS80}, 
\begin{eqnarray}
|{\rm HFB}\rangle&=&
\prod_{p>0}(u_p+v_pa^\dagger_pa^\dagger_{\bar{p}})|0\rangle, 
\label{HFBcan}
\\
&\propto& 
\exp\left(\sum_{p>0}\frac{v_p}{u_p}\,a^\dagger_pa^\dagger_{\bar{p}}\right)|0\rangle, 
\label{HFBwf2}
\end{eqnarray}
where $u_p=\sqrt{1-v_p^2}$ and $\bar{p}$ is the time-reversed state of
$p$. Since the creation operator for the canonical basis,
$a^\dagger_p$, is related to the creation operator in the coordinate
space, $a^\dagger_{\vec{r}\sigma}$, as 
\begin{equation}
a^\dagger_p=\int d\vec{r}\sum_\sigma\,\psi_p(\vec{r},\sigma)a^\dagger_{\vec{r}\sigma},
\end{equation}
Eq. (\ref{HFBwf2}) is transformed to \cite{DNW96}, 
\begin{equation}
|{\rm HFB}\rangle
\propto
\exp\left(-\frac{1}{2}\int d\vec{r}d\vec{r}'\sum_{\sigma,\sigma'}
Z(\vec{r}\sigma,\vec{r}'\sigma')\,
a^\dagger_{\vec{r}\sigma}a^\dagger_{\vec{r}'\sigma'}
\right)|0\rangle, 
\end{equation}
with 
\begin{equation}
Z(\vec{r}\sigma,\vec{r}'\sigma')
=-2\sum_{p>0}
\frac{v_p}{u_p}\,\psi_p(\vec{r},\sigma)\psi_{\bar{p}}(\vec{r}',\sigma').
\end{equation}
To evaluate the four-particle density, (\ref{rho4}), 
we first perform the particle number projection onto the HFB state, 
\begin{equation}
\hat{P}_N|{\rm HFB}\rangle \propto
\left(-\frac{1}{2}\int d\vec{r}d\vec{r}'\sum_{\sigma,\sigma'}
Z(\vec{r}\sigma,\vec{r}'\sigma')\,
a^\dagger_{\vec{r}\sigma}a^\dagger_{\vec{r}'\sigma'}
\right)^2|0\rangle.
\label{PNP}
\end{equation}
The four-particle density is then obtained as 
\begin{equation}
\rho_4(x_1,x_2,x_3,x_4) \propto |f(x_1,x_2,x_3,x_4)|^2,
\end{equation}
with 
\begin{eqnarray}
&&f(x_1,x_2,x_3,x_4)\nonumber \\
&=&
(Z(x_1,x_2)-Z(x_2,x_1))(Z(x_3,x_4)-Z(x_4,x_3)) \nonumber \\
&+&(Z(x_1,x_3)-Z(x_3,x_1))(Z(x_4,x_2)-Z(x_2,x_4)) \nonumber \\
&+&(Z(x_1,x_4)-Z(x_4,x_1))(Z(x_2,x_3)-Z(x_3,x_2)).
\end{eqnarray}

\subsection{Probability for shell model configurations}

Using the canonical basis representation of the HFB state,
(\ref{HFBcan}), 
one can also calculate the probability for a shell model configuration, 
[$(k\bar{k})(k'\bar{k}')$], 
for the four-particle systems 
when $k$ and $k'$ represent the canonical
basis states.  
It is given by  
\begin{equation}
P_{k^2k'^2}=
\frac{1}{{\cal N}}
|\langle 
(k\bar{k})(k'\bar{k}')|{\rm HFB}\rangle|^2 
=\frac{1}{{\cal N}}
v_k^2v_{k'}^2\prod_{p\neq k,k'(>0)} u_p^2,
\label{Pshell}
\end{equation}
where the normalization factor ${\cal N}$ reads
\begin{equation}
{\cal N}=
\langle {\rm HFB}|\hat{P}_{N=4}|{\rm HFB}\rangle 
=\frac{1}{2\pi}\int^{2\pi}_0d\phi 
e^{-4i\phi}
\prod_{p>0}(u_p^2+e^{2i\phi}v_p^2).
\end{equation}
Here, we have used the explicit form of the number projection operator
\cite{RS80}, 
\begin{equation}
\hat{P}_N= 
\frac{1}{2\pi}\int^{2\pi}_0d\phi 
e^{i\phi(\hat{N}-N)}.
\end{equation}

If the angular momentum components are explicitly expressed, the
probability (\ref{Pshell}) reads
\begin{equation}
P_{(lj)^4}=\frac{1}{{\cal N}} \cdot 
\left(
\begin{array}{c}
\Omega_j\\
2 
\end{array}
\right)
(v_{lj}^2)^2(u_{lj}^2)^{\Omega_j-2}
\prod_{l'j'\neq lj}
(u_{l'j'}^2)^{\Omega_{j'}},
\end{equation}
for the $(lj)^4$ configuration, while 
\begin{eqnarray}
P_{(lj)^2(l'j')^2}&=&\frac{1}{{\cal N}}\cdot 
\Omega_j\Omega_{j'}\,
v_{lj}^2(u_{lj}^2)^{\Omega_j-1}
v_{l'j'}^2(u_{l'j'}^2)^{\Omega_{j'}-1} \nonumber \\
&&\times \prod_{l''j''\neq lj,l'j'}
(u_{l''j''}^2)^{\Omega_{j''}},
\end{eqnarray}
for the $(lj)^2(l'j')^2$ configuration with $lj\neq l'j'$. 
In these equations, $\Omega_j=(2j+1)/2$ is the pair degeneracy for the 
angular momentum $j$ state. 

\section{Applicability of HFB method: test on $^6$He nucleus}

Let us now numerically solve the HFB equations and discuss the spatial
structure of neutron-rich nuclei. Before we do this, however, we first 
examine the applicability of the HFB method by applying it to the 
three-body model of $^6$He nucleus \cite{HS05,EBH99}. This model 
has been solved exactly by
diagonalizing the Hamiltonian matrix. A comparison of the HFB
solution with the exact result for this model 
will provide an idea on whether the HFB method
is good enough to discuss the dineutron correlation in neutron-rich
nuclei. 

To this end, 
we use the same neutron-core potential, $V_{nC}$, and the
same pairing interaction $v_{nn}$ as in Refs. \cite{HS05,EBH99}. 
The neutron-core potential is taken as a Woods-Saxon form. 
Since we use the renormalized mass $m$ (see Eq. (\ref{H5body})) instead of the reduced mass, 
we multiply the factor $(A-1)/A\cdot A_C/(A_C+1)$, where $A_C$ is the
mass number of the core nucleus, following the prescription given in
Ref. \cite{EBH99}. The pairing interaction is given as 
\begin{equation}
v_{nn}(\vec{r},\vec{r}')=\delta(\vec{r}-\vec{r}')
\left(v_0+\frac{v_\rho}{1+\exp[(r-R_\rho)/a_\rho]}\right).
\label{vnn3body}
\end{equation}
We multiply an overall scaling factor to this interaction so that the
two-neutron separation energy of $^6$He is reproduced with the HFB
method. We use the same value for all the other parameters as in
Ref. \cite{HS05}. 
Note that the last term in  Eq. (\ref{hmf}) disappears for the 
pairing interaction given by Eq. (\ref{vnn3body}), 
since the interaction does not depend explicitly on
the density, but the density dependence is parameterized by the Fermi
function. 

\begin{figure}[htb]
\includegraphics[scale=1.1]{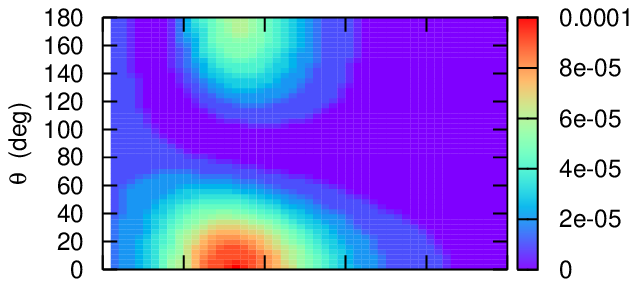}\\
\vspace{-1.5cm}
\includegraphics[scale=1.1]{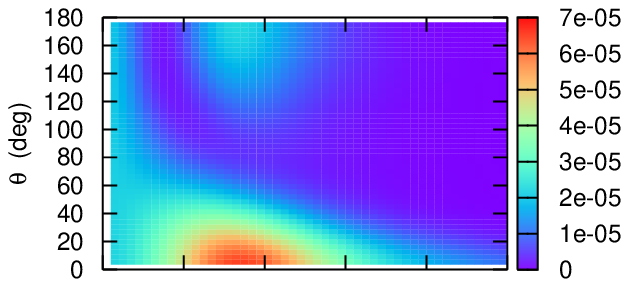}\\
\vspace{-1.5cm}
\includegraphics[scale=1.1]{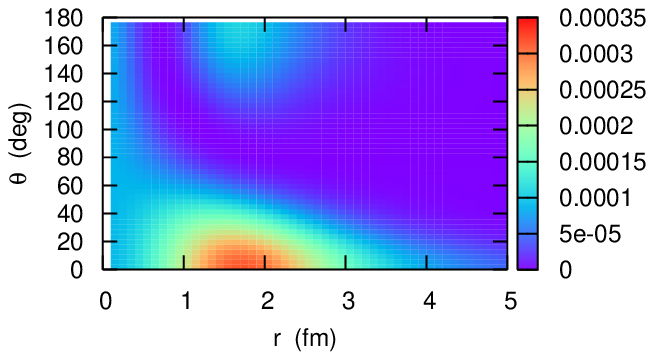}
\caption{(Color online) 
The $S=0$ component of the two-particle density for $^6$He 
as a function of $r_1=r_2=r$ and the angle between 
the valence neutrons, $\theta$. 
The top panel shows the exact solution of the three-body model, 
while the middle panel is obtained with the HFB method. 
The bottom panel shows the result of the HFB + particle number
projection. 
}
\end{figure}

Figure 1 shows the two-particle density for $^6$He in the $S=0$
channel, that is, $\rho_2(\vec{r}_1\uparrow,\vec{r}_2\downarrow)$. 
As we have done in Ref. \cite{HS05}, 
we set $r_1=r_2\equiv r$ and plot the density as a function of $r$ and
the relative angle between the spin up and down neutrons, $\theta$. 
Figure 2 shows the same two-particle density, but we multiply the factor
$8\pi^2r^4\sin\theta$ \cite{HS05}. 
The top panels in these figures show the exact solution of the
three-body Hamiltonian \cite{HS05}, while the middle panels are for
the HFB results. 
One can clearly see that the localization of the two-particle density around
$\theta\sim 0$ 
in the three-body model
is well reproduced by the HFB method, although the
HFB density has a somewhat longer tail and the density around 
$\theta\sim \pi$ is largely suppressed. 
The localization of the two-particle density is nothing but the 
manifestation of the strong di-neutron correlation in a halo nucleus 
 $^6$He.  The longer tail of the HFB density may be due to the 
 asymptotic behavior of the pair density    
 $ \tilde{\rho}_{v}(\vec{r}_1\uparrow,\vec{r}_2\downarrow)$
 in Eq. (\ref{rho_2}), which is different from that of the normal density 
  ${\rho}_{v}(\vec{r}_1\uparrow,\vec{r}_2\downarrow)$ \cite{DNW96}.
The similarity between the exact and the HFB results 
for the two-particle density   
is rather
striking, and it is clear that the HFB method can be utilized to
discuss, at least qualitatively, the strong dineutron correlation in
neutron-rich nuclei. 

\begin{figure}[htb]
\includegraphics[scale=1.1]{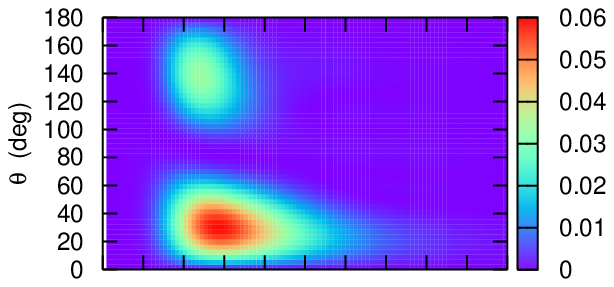}\\
\vspace{-1.5cm}
\includegraphics[scale=1.1]{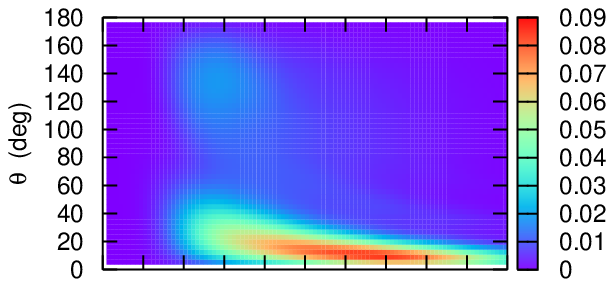}\\
\vspace{-1.5cm}
\includegraphics[scale=1.1]{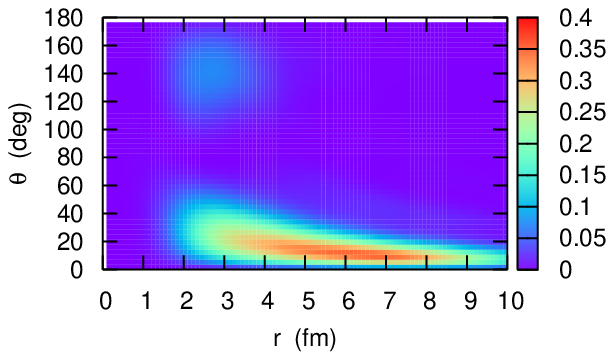}
\caption{(Color online) 
Same as Fig.1, but with a multiplicative factor of $8\pi^2 r^4\sin\theta$. 
}
\end{figure}

We also study the effect of particle number projection on the
two-particle density. The bottom panels in Figs. 1 and 2 are obtained
by taking the number projection onto the HFB ground state 
(with the variation before projection (VBP) scheme \cite{HRB02}) 
in a similar way as in Eq. (\ref{PNP}). 
The two-particle density thus obtained is not normalized and 
the scale is different between the middle and the bottom panels. 
However, we can see that the dependence of the two-particle density on
$r$ and $\theta$ is almost the same between the two panels. Therefore,
we conclude that 
the effect of number projection on the two-particle density is 
rather small as far as the two-particle density is concerned, 
although the projection might still affect the
density if the variation after projection (VAP) scheme is employed. 

\begin{table}[hbt]
\caption{
Comparison of the exact and the HFB results for the occupation
probabilities in the ground state of $^6$He. 
}
\begin{center}
\begin{tabular}{c|c|c}
\hline
\hline
configuration & exact \cite{HS05} & HFB \\
\hline
(s$_{1/2})^2$ & 3.04 \% & 7.25 \% \\
(p$_{1/2})^2$ & 4.85 \% & 9.21 \% \\
(p$_{3/2})^2$ & 83.0 \% & 57.4 \% \\
(d$_{3/2})^2$ & 1.47 \% & 3.86 \% \\
(d$_{5/2})^2$ & 6.11 \% & 6.85 \% \\
(f$_{5/2})^2$ & 0.035 \% & 2.30 \% \\
(f$_{7/2})^2$ & 0.075 \% & 3.33  \% \\
\hline
\hline
\end{tabular}
\end{center}
\end{table}

Table 1 summarizes the occupation probabilities for the $^6$He
nucleus. Although the absolute value is somewhat smaller, the HFB well
reproduces the dominance of the (p$_{3/2})^2$ configuration in the
ground state wave function. 
Again, the HFB method provides a good estimate of the ground state
properties of neutron-rich nuclei even when the number of particle is as small as two.

\section{Dineutron correlation in $^8$He and $^{18}$C}

We now solve the HFB equations for the $^8$He and $^{18}$C nuclei. 
We use the same neutron-core potential, 
$V_{nC}$, for $^8$He as in Refs. \cite{HS05,EBH99}, while we use the
set D in Refs. \cite{HS07,VM95} for the $^{18}$C nucleus. 
As in the previous section, these potentials are scaled by a 
factor of $(A-1)/A\cdot A_C/(A_C+1)$. 
For the core density, $\rho_C$, 
we use those in Refs. \cite{ZKS94,PZV00}. 
We determine the strength of the pairing interaction 
so that the experimental ground state energy relative to the core+4$n$
threshold, $E=-3.112$ MeV for $^8$He and $-$10.385 MeV for $^{18}$C, 
is reproduced with $\rho_0$=0.32 fm$^{-3}$ ({\it i.e.,} the mixed pairing
interaction\cite{DN02,DNS02}). With the cut-off energy of
$\epsilon_{\rm cut}+\lambda$=40 MeV in the single-particle
space, this leads to 
$V_0=-$502 MeV fm$^3$ for $^8$He and 
$V_0=-$538 MeV fm$^3$ for $^{18}$C. 

\begin{table}[hbt]
\caption{
The results of the HFB calculation for the Fermi energy $\lambda$ and
the root-mean-square (rms) radius, $r_{\rm rms}$, for 
the $^8$He and $^{18}$C nuclei. 
}
\begin{center}
\begin{tabular}{c|cccc}
\hline
\hline
nucleus & $E_{g.s.}$ (MeV) & $\lambda$ (MeV) & $r_{\rm rms}$ (fm) &  
$r^{\rm (exp)}_{\rm rms}$ (fm)  \\
\hline
$^8$He & $-3.112$ & $-0.0715$ & 3.23 & 2.49 $\pm$ 0.04 \cite{THKSST92} \\
$^{18}$C & $-10.514$ & $-2.522$ & 2.92 & 2.90 $\pm$ 0.19 \cite{LBG90}\\
\hline
\hline
\end{tabular}
\end{center}
\end{table}

The results of the HFB calculation are summarized in Table II. 
Although our purpose in this paper is not to reproduce the
experimental data, but to discuss qualitatively the dineutron
correlation in $^8$He and $^{18}$C, we notice that the
root-mean-square radius for the $^{18}$C nucleus is well reproduced
with the present calculation. 
The mean-field potential in Eq. (\ref{hmf}) is shown in Fig. 
3, in which the dashed and the solid lines correspond to the
neutron-core potential, $V_{nC}$, and the total mean-field potential,
respectively. 
The difference between the two potentials 
originates from the effect of pairing correlations among 
the valence particles on the mean field potential. As a consequence, 
 the mean-field potential 
for $^8$He posses one bound single-particle state  while 
the neutron-$^4$He potential $V_{nC}$
alone does not hold any bound single-particle state, 
  reflecting the Borromean nature
of the $^6$He nucleus \cite{HS05,EBH99}.  
For the $^{18}$C nucleus, the same effect shifts the single-particle
energy from $-1.072$ 
to $-1.768$ MeV for the 2s$_{1/2}$ state and from $-0.414$ 
to $-1.664$ MeV for 1d$_{5/2}$. 

\begin{figure}[hbt]
\includegraphics[scale=0.4,clip]{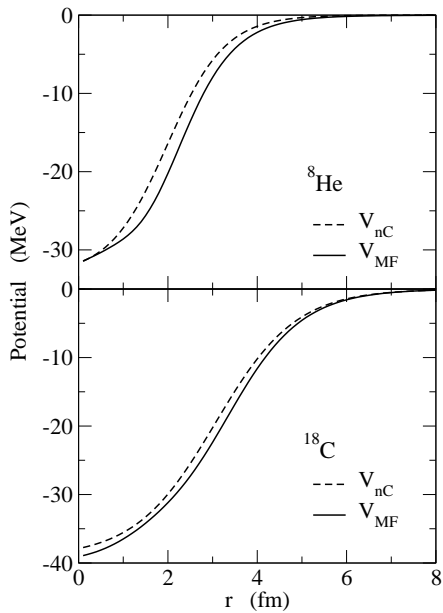}
\caption{
The mean-field potential for $^8$He (the upper panel) and 
for $^{18}$C (the lower panel). 
The dashed line shows the neutron-core potential, $V_{nC}$, while 
the solid line is for the total mean-field potential in the solution
of the HFB equations. 
}
\end{figure}

\begin{table}[hbt]
\caption{
Probability of a few shell model configurations in the HFB ground 
state wave function for $^8$He and $^{18}$C. 
}
\begin{center}
\begin{tabular}{c|cc}
\hline
\hline
nucleus & configuration & probability (\%) \\
\hline
$^8$He & [(1p$_{3/2})^4$] & 34.9 \\
       & [(1p$_{3/2})^2$(p$_{1/2})^2$] & 23.7 \\
       & [(p$_{3/2})^2$(d$_{5/2})^2$] & 10.7 \\
       & [(s$_{1/2})^2$(p$_{3/2})^2$] & 7.8 \\
\hline
$^{18}$C & [(1d$_{5/2})^4$] & 32.2 \\
        & [(1d$_{5/2})^2$(2s$_{1/2})^2$] & 26.2 \\
        & [(1d$_{5/2})^2$(d$_{3/2})^2$] & 11.8 \\
       & [(d$_{5/2})^2$(f$_{7/2})^2$] & 7.17 \\
\hline
\hline
\end{tabular}
\end{center}
\end{table}

The probability for a few single-particle components in the ground
state wave function is listed in Table III.  
For the $^8$He nucleus, although the largest probability is found 
in the [(1p$_{3/2})^4$] configuration, the other components also
have an appreciable probability. Therefore, this nucleus largely
deviates from the pure 
[(1p$_{3/2})^4$] configuration, 
in accordance with the recent experimental finding with the 
$^8$He($p,t)^6$He reaction \cite{KSL07}. 
For the $^{18}$C nucleus, the ground state wave function mainly
consists of the 
[(1d$_{5/2})^4$] and the 
[(1d$_{5/2})^2$(2s$_{1/2})^2$] configurations,
while the [(1d$_{5/2})^2$(d$_{3/2})^2$] and the 
[(d$_{5/2})^2$(f$_{7/2})^2$] 
 configurations  are also appreciable in 
 Table III. 

The top panel in Figs. 4 and 5 shows the two-particle
density, 
$\rho_2(r,\hat{\vec{r}}=0,\uparrow;r,\hat{\vec{r}},\downarrow)$, for
$^8$He and $^{18}$C, respectively. 
The middle panels in these figures show the same two-particle density, 
but with the multiplicative factor of $8\pi^2r^4\sin\theta$. 
For both the 
$^8$He and $^{18}$C nuclei, one clearly finds a strong concentration of
two-particle density around $\theta\sim 0$ at around the nuclear
surface. This is similar to what has been found in the Borromean
nuclei, $^{11}$Li and $^6$He \cite{HS05,EBH99} (see also Figs. 1 and
2), and indicates clearly the strong dineutron correlation 
\cite{HSCP07,MMS05,PSS07} in these nuclei. 

\begin{figure}[t]
\includegraphics[scale=1.1]{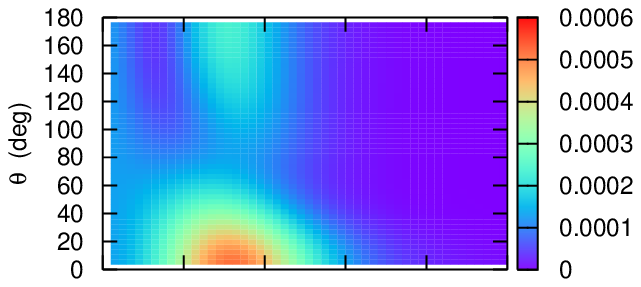}\\
\vspace{-1.5cm}
\includegraphics[scale=1.1]{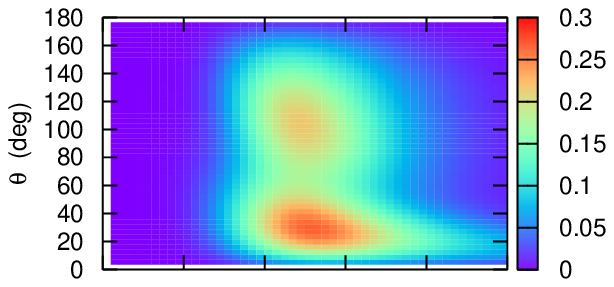}\\
\vspace{-1.5cm}
\includegraphics[scale=1.1]{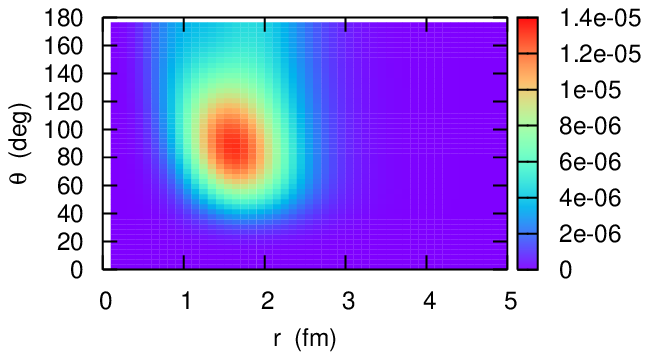}
\caption{(Color online) 
The two-particle density, 
$\rho_2(r,\hat{\vec{r}}=0,\uparrow;r,\hat{\vec{r}},\downarrow)$, for 
the $^8$He nucleus as a function of $r_1=r_2=r$ and the relative angle
$\theta$ between a spin-up and a spin-down neutrons (the top panel). 
The middle panel shows the same two-particle density multiplied by a
factor $8\pi^2r^4\sin\theta$, while the bottom panel is for the
four-particle 
density for the dineutron-dineutron configuration. 
}
\end{figure}

\begin{figure}[htb]
\includegraphics[scale=1.1]{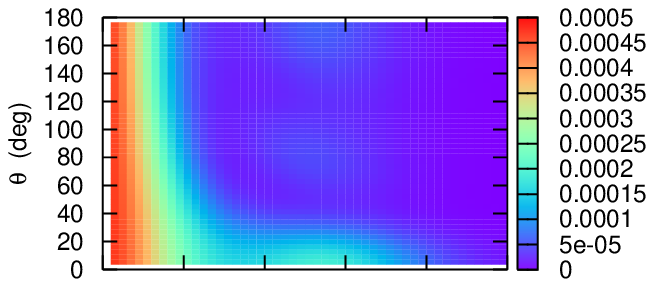}\\
\vspace{-1.5cm}
\includegraphics[scale=1.1]{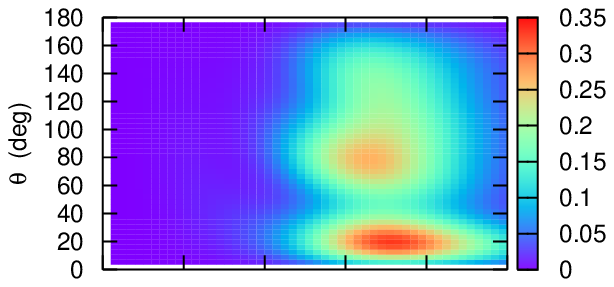}\\
\vspace{-1.5cm}
\includegraphics[scale=1.1]{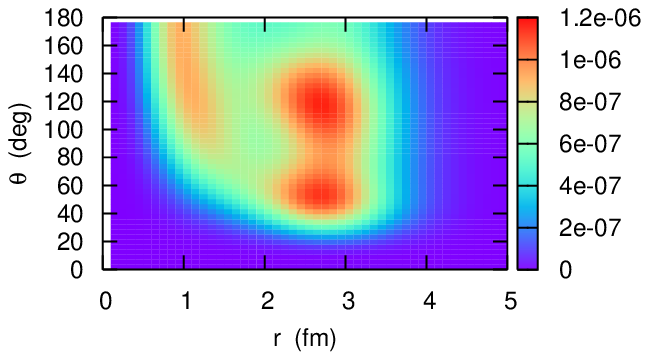}
\caption{(Color online) 
Same as Fig. 4, but for $^{18}$C. 
}
\end{figure}

Since the strong dineutron structure is apparent for a spin-up and a
spin-down neutrons in these nuclei, 
we next plot the four-particle density for the two-dineutron configuration, 
that is, the four-particle density with
$x_1=(r,\hat{\vec{r}}=0,\uparrow),~x_2=(r,\hat{\vec{r}}=0,\downarrow), 
~x_3=(r,\hat{\vec{r}},\uparrow)$, and 
$x_4=(r,\hat{\vec{r}},\downarrow)$ in Eq. (\ref{rho4}). 
This is plotted in the bottom panels in Figs. 4 and 5 for 
$^8$He and $^{18}$C, respectively. 
For the $^8$He nucleus, the four-particle density for the 
dineutron-dineutron configuration has a peak around $\theta\sim \pi/2$. 
A similar result has been obtained with a three-body model calculation
with the dineutron clusters, that is, $\alpha+n^2+n^2$ \cite{NVC01}. 
The peak around $\theta\sim \pi/2$ arises from the main component of the wave
function, that is, the [(1p$_{3/2})^4$] configuration, for which the
four-particle density is proportional to $\sin^4\theta\propto
|Y_{11}|^4$. For the $^{18}$C nucleus, the four-particle density has
two peaks, one around $\theta\sim 54$ deg. and the other around 
$\theta\sim 118$ degree. 
This can again be understood in terms of the 
[(1d$_{5/2})^4$] configuration, for which the four-particle density 
is proportional to $(3|Y_{22}|^2+2|Y_{21}|^2)^2$. 

To demonstrate more clearly the similarity between the four-particle
density to that for the main components, the bottom panels of Figs. 
6 and 7 show the four-particle density for the 
[(1p$_{3/2})^4$] and [(1d$_{5/2})^4$] configurations in the 
neutron-core potential $V_{nC}$, respectively. 
To this end, we adjust the depth of the neutron-core potential so that 
the energy of the 1p$_{3/2}$ and 1d$_{5/2}$ states is 
a quarter the energy of $^8$He and $^{18}$C, respectively. 
The similarity between the four-particle density for the correlated wave
functions (Figs. 4 and 5) and that for the uncorrelated wave functions
(Figs. 6 and 7) is apparent.
This is a natural consequence of a short range nature of 
  nuclear interaction:  
the two neutrons with the same spin have to be far apart in space
due to the Pauli principle and thus their distance is likely larger than 
the range of the nuclear interaction. As a consequence, the
interaction between the two 
dineutrons becomes weak, despite that the
two neutrons  
in the {\it same} dineutron having different spins 
  strongly interact with each other. 
From this consideration, we conclude that the 
two dineutrons are moving rather freely  in the core+4n nuclei 
respecting  solely the Pauli principle.

Note that the pairing interaction yet plays an essential role in the
two-particle density. Without the pairing correlation, the
two-particle density for the uncorrelated wave functions has 
symmetric bumps both around $\theta\sim 0$ and $\theta\sim\pi$, as is
shown in the upper panel in Figs. 6 and 7 (see also the middle panels,
that show the two-particle density with the weight of
8$\pi^2r^4\sin\theta$). The pairing correlation mixes several angular
momentum components in the ground state wave function, eliminating essentially the bump around
$\theta\sim\pi$.
The configuration mixing of  different parity states seen in Table III is  
 essential to have the di-neutron peak in the middle panel of Figs. 4 and 5. 

\begin{figure}[htb]
\includegraphics[scale=1.1]{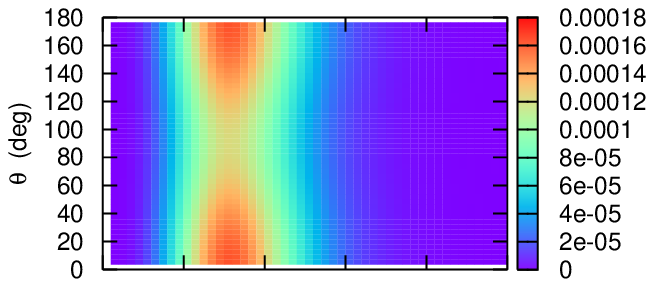}\\
\vspace{-1.5cm}
\includegraphics[scale=1.1]{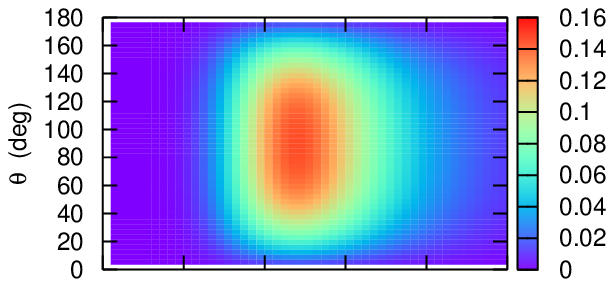}\\
\vspace{-1.5cm}
\includegraphics[scale=1.1]{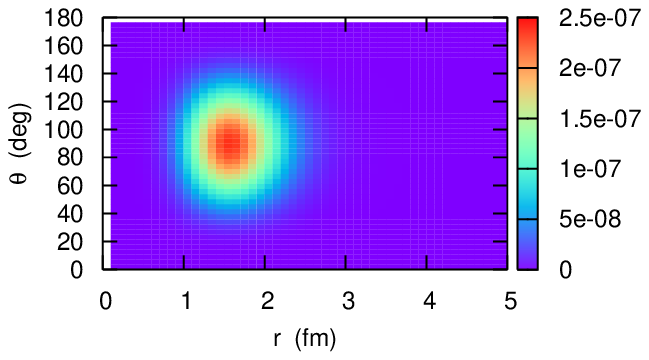}
\caption{(Color online) 
Same as Fig.4, but for the uncorrelated 
[(1p$_{3/2})^4$] configuration for the $^8$He nucleus. 
}
\end{figure}

\begin{figure}[htb]
\includegraphics[scale=1.1]{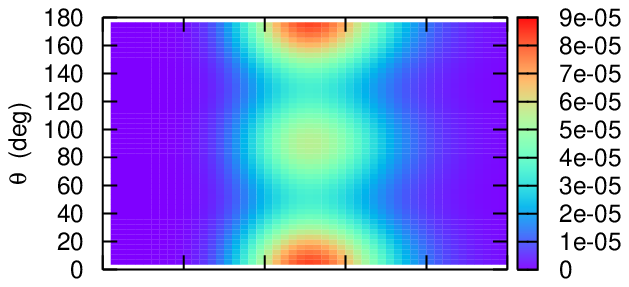}\\
\vspace{-1.5cm}
\includegraphics[scale=1.1]{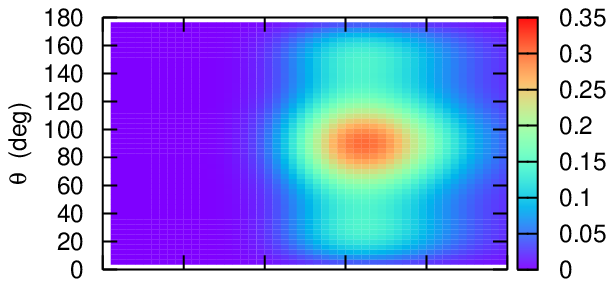}\\
\vspace{-1.5cm}
\includegraphics[scale=1.1]{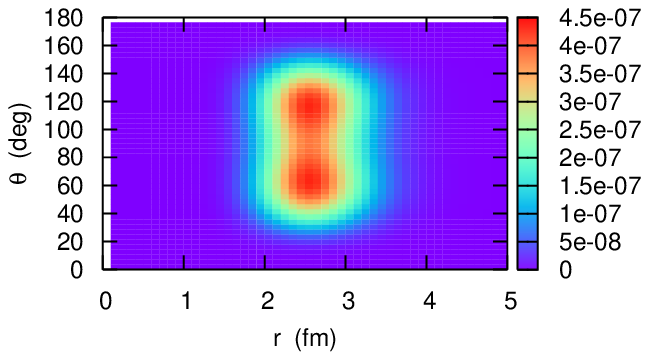}
\caption{(Color online) 
Same as Fig.5, but for the uncorrelated 
[(1d$_{5/2})^4$] configuration for the $^{18}$C nucleus. 
}
\end{figure}

Another way to investigate the four-particle density is to plot the density
distribution of the second dineutron when the first dineutron is put 
on the $z$-axis \cite{MMS05}, rather than assuming that the distance from the core
nucleus is the same between the two dineutrons. 
The top, middle, and bottom panels of Fig. 8 show the
four-particle density of $^8$He for the dineutron-dineutron configuration when
the first dineutron is at $z$=1.5, 2.5, and 3.5 fm, respectively. The
pairing correlation is taken into account in the plot. The
same plot for the $^{18}$C nucleus is shown in Fig. 9. 
These figures demonstrate that the distance of the second dineutron
from the core, $r_2$, increases as the distance of the first
dineutron, $r_1$, increases, tending to $r_1\sim r_2$. 
The angular distribution of the second dineutron, on the other hand,
is almost independent of the position of the first dineutron. 
This behaviour is 
consistent 
with the four-particle density 
shown in the bottom panels of Figs. 4 and 5. 

\begin{figure}[htb]
\includegraphics[scale=1.1]{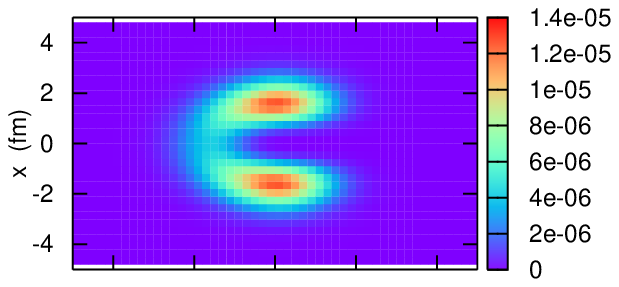}\\
\vspace{-1.5cm}
\includegraphics[scale=1.1]{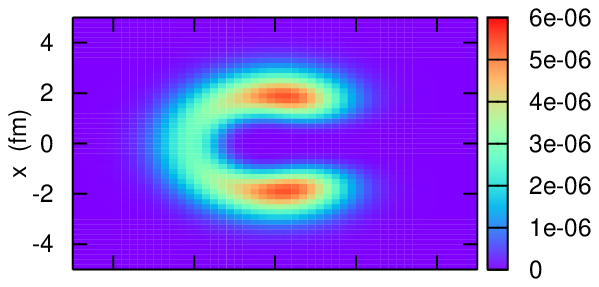}\\
\vspace{-1.5cm}
\includegraphics[scale=1.1]{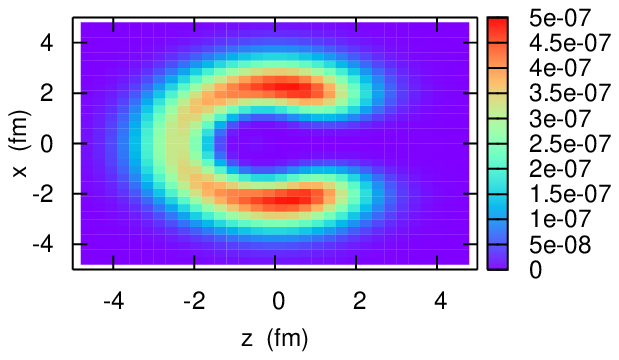}
\caption{(Color online) 
The four-particle density of $^8$He for the dineutron-dineutron
configuration when the first dineutron is on the $z$-axis. 
The top, middle, and bottom panels correspond to the cases where the 
first dineutron is at $z$=1.5, 2.5, and 3.5 fm, respectively. 
}
\end{figure}

\begin{figure}[htb]
\includegraphics[scale=1.1]{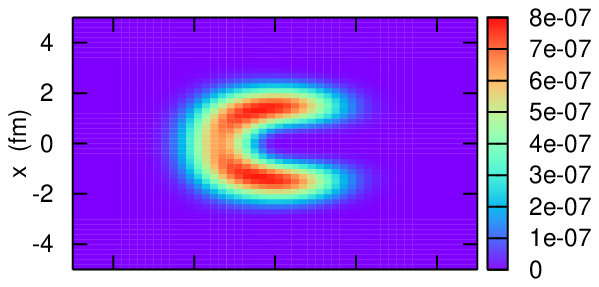}\\
\vspace{-1.5cm}
\includegraphics[scale=1.1]{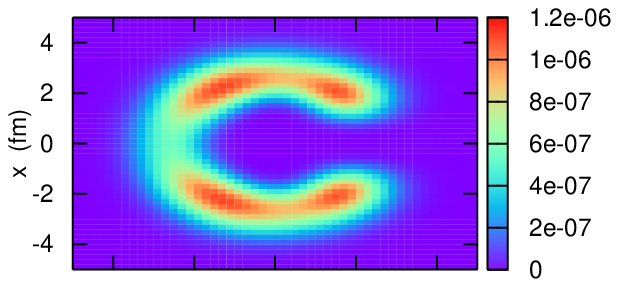}\\
\vspace{-1.5cm}
\includegraphics[scale=1.1]{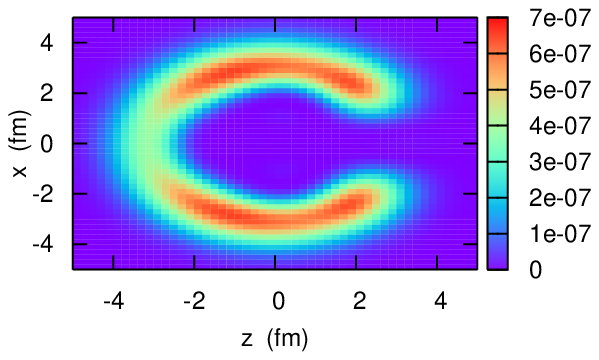}
\caption{(Color online) 
Same as Fig. 8, but for $^{18}$C. 
}
\end{figure}

\section {Summary}

We have discussed the dineutron structure in 
the 2$n$ halo nucleus $^6$He as well as in the core+4$n$ nuclei, 
$^8$He and $^{18}$C. For this purpose, we 
employed the density-dependent contact interaction among the valence
neutrons, and diagonalized the 
core+$x$n Hamiltonian ($x$=2 for $^6$He and $x=4$ for $^8$He and
$^{18}$C) with the Hartree-Fock-Bogoliubov (HFB) method. 
From the comparison with the exact solution of the three-body
Hamiltonian for the $^6$He nucleus, we found that the HFB method is
satisfactory enough to discuss the spatial structure of the valence
neutrons. For the $^8$He and $^{18}$C nuclei, we investigated both the
two- and the four-particle densities. We showed that the two-particle density takes
the largest value when the spin-up and the spin-down neutrons are at
the same position, that is nothing but the manifestation of the strong dineutron
correlation. 
With this result in mind, we particularly discussed the four-particle
density for the dineutron-dineutron configuration. 
We found that two dineutrons weakly interact with each other,
simply respecting the Pauli principle. 
The four-particle density of the HFB calculation 
  in fact resembles to that for the uncorrelated wave
functions. 
This result is entirely due to a short range nature of nuclear interaction.  
Namely, the two neutrons with the same spin have to be far apart in space
due to the Pauli principle and thus their distance is likely larger than 
the range of the nuclear interaction. 
As a consequence, the
interaction between the two 
dineutrons becomes weak, while  the
two neutrons in the {\it same} dineutron strongly interact with each other.

We have also discussed the probability for the single-particle
configurations in the ground state wave function. Our HFB calculations
indicate that the $^8$He nucleus consists of the 
[(1p$_{3/2})^4$] configuration by 34.9\% and 
of the
[(1p$_{3/2})^2$(p$_{1/2})^2$] configuration by 23.7\%, while the $^{18}$C
nucleus consists of the 
[(1d$_{5/2})^4$] and 
the [(1d$_{5/2})^2$(2s$_{1/2})^2$] configurations 
by 32.2\% and 26.2\%, respectively. 
The result for the $^8$He nucleus is consistent with the recent
experimental finding with the two-neutron transfer reaction, 
$^8$He($p,t)^6$He, that indicates 
an appreciable mixture
of the configurations other than [(1p$_{3/2})^4$], {\it e.g.,} 
[(1p$_{3/2})^2$(p$_{1/2})^2$], in the ground state of
$^8$He. 
It would be interesting to analyse the 
experimental data for the $^8$He($p,t)^6$He reaction with the 
wave function obtained in this paper. 
This will be a topic for a future publication. 

\begin{acknowledgments}
We thank J. Dobaczewski for useful discussions. 
This work was supported by the Japanese
Ministry of Education, Culture, Sports, Science and Technology
by Grant-in-Aid for Scientific Research under
the program number 19740115. 
\end{acknowledgments}

\end{document}